\documentclass[aps,prl,twocolumn]{revtex4}
\usepackage{graphicx}
\begin{document}
\title{Excitons in carbon  nanotubes: 
an {\it ab initio}  symmetry-based approach}

\author{Eric Chang, Giovanni Bussi, Alice Ruini, and Elisa Molinari}
\address{INFM National Center on nanoStructures and bioSystems
at Surfaces (S$^3$) and Dipartimento di Fisica, \\
Universit\`{a}
di Modena e Reggio Emilia, Via Campi 213/A, 41100 Modena, Italy \\}
\date{\today}
\begin{abstract}
The optical absorption spectrum of the carbon
$(4,2)$ nanotube is computed using an {\it ab initio}
many-body approach which takes into account excitonic effects.
We develop a new method involving a local basis set
which is symmetric with 
respect to the screw symmetry of the tube. Such a method has  
the advantages of scaling faster    
than plane-wave methods and  
allowing for a precise determination
of the symmetry character of the single particle states,
two-particle excitations, and selection rules.
The binding energy of the 
lowest, optically active states is approximately 0.8~eV.
The corresponding exciton wavefunctions are delocalized along
the circumference of the tube and localized in the direction of
the tube axis.
\end{abstract}
\pacs{71.15.Qe,78.67.Ch,78.20.Bh}
\maketitle

A wealth of extraordinary results concerning the 
mechanical and electrical
properties of carbon nanotubes (NTs) have been reported in the last 
few years~\cite{sait+98book:note}. 
Until recently, however, their optical properties have not received
the same attention: experimental work has often been hindered by low
emission efficiency, and the interpretation has been 
complicated by the fact
that tubes of different species and orientation are normally mixed
together in the
same sample, making it difficult to assign the measured spectra
to a single NT species.

Very recent experiments have indicated that these 
limitations can be overcome.
Improved optical efficiency has been obtained by isolating NTs in porous
materials~\cite{li+01prl}, in solution~\cite{ocon+03sci}, or on patterned
substrates~\cite{leve+03prl}. It was possible
to assign optical spectra to specific
NTs via their characteristic vibrations in resonant 
Raman~\cite{bach+03sci} 
or in near-field experiments~\cite{hart+03prl}. The
observation of electrically induced optical emission from a carbon NT 
FET~\cite{mise+03sci} has paved 
the way for a new class of single-molecule
experiments and devices. These advances therefore establish optical
spectroscopies as powerful characterization tools for NTs. 
Nanotubes also hold 
great promise for novel nanoscale opto-electronic and photonic
applications~\cite{mise+03sci}, because
the
optical gap of NTs spans a very large frequency range
which overlaps with the range 
of interest in the field of telecommunications. 

In spite of such fervent interest in this subject,
 the fundamental nature of optical
excitations of NTs is not yet understood. The possible relevance of
excitonic effects in these systems 
was pointed out in a pioneering paper by
Ando~\cite{ando97jpsj}. In general, it is 
well known that the electron-hole
interaction plays a crucial role in one-dimensional systems, not
only in the ideal case~\cite{loud59ajp}, 
but also in realistic systems such
as semiconductor quantum wires~\cite{ross-moli96prl} or polymer 
chains~\cite{ruin+02prl} 
where excitons dominate the optical spectra. The binding energies are
however very sensitive to the spatial extent of the single-particle
wavefunctions and to 
the (anisotropic) dielectric screening~\cite{foot1}.  In
the case of NTs, one might expect these quantities to be sensitive
to size and geometry. Moreover, given the peculiar nature of the 
electronic states,
even the smaller diameter NTs 
cannot be regarded as pure one-dimensional systems. 
It is therefore extremely difficult to
estimate the actual 
relevance of excitonic effects without a realistic
calculation of their optical excitations.

In this Letter, we present an efficient {\it ab initio} approach 
to compute the spectra, 
exciton energies and wavefunctions of carbon NTs. Such an approach
includes the full three-dimensional dependence of the electron
states and interactions.  Our formalism exploits 
a fully symmetrized Gaussian basis set and allows us not 
only to reduce the system size and computation time considerably, 
but also to profit from the selection rules involving
a new quantum number which comes into play in the
symmetry characterization of the problem. 

As a preliminary step, a calculation based on the
density-functional theory within the local density   
approximation~\cite{kohn-sham65pr} is performed with a 
 plane-wave basis, pseudopotentials
and supercells~\cite{pwscf} and 
the band states $\psi_{n{k}}({\bf G})$ are calculated
in Fourier space. These states are then projected onto a
set of basis functions which are  symmetrized  sums of
Gaussian orbitals centered on the atoms. The basis is constructed
to be simultaneous eigenstates of the commuting
operators  $\hat{G}_1$ and $\hat{G}_2$. 
$\hat{G}_1$ represents a discrete translation of the tube 
in the $z$-direction by the length of the cell $T$.
$\hat{G}_2$ is the screw-symmetry operation
consisting of a  combined rotation and translation: a rotation
by an angle of $2 \pi/N$, where $N$ is the number
of hexagons in the supercell; and a translation along
the $z$-direction by $(M/N)T$, where $M$ depends on the geometry
and size of 
the tube (for details see Ref.~\cite{sait+98book:note}). 
The basis functions have the form
\begin{equation}
\chi_{\sigma nlm}^{{\bf q} h}({\bf r}) =
\sum_{j_R = 1}^{N}
\sum_{j_T = 1}^{N_{cell}}
f_{\bf q}^{j_T} g_{{\bf q} h}^{j_R}
[\hat{G}_2^{j_R} \hat{G}_1^{j_T} \phi_{nlm}({\bf
 r} - \tau_{\sigma})].
\end{equation}
The functions 
$\lbrace \phi_{nlm}({\bf r}) \rbrace$ are
Gaussians defined as
$\phi_{nlm}({\bf r}) = r^l \exp({{-r^2} /
{a_n^2} })
                      Y_{l m} ({\bf \hat{r}})$. 
They have been used with considerable
success in similar 
calculations for extended systems~\cite{rohl+95prb}.
$N$ is the number of hexagons per unit cell. $\sigma = 1,2$
labels one of the two inequivalent atoms in the unit cell. 
$f_{\bf q} = e^{2 \pi i q }$ and
$g_{{\bf q}h} = e^{2 \pi i (Mq+h)/N}$ are the
eigenvalues of the operators $\hat{G}_1$ and $\hat{G}_2$,
respectively. $h$ is a new quantum number, analogous to the azimuthal
quantum number $m$, which ranges from $1$ to $N$.

The quantities needed for the calculation of both the random phase 
approximation (RPA) dielectric function~\cite{adle62pr}
and the Bethe Salpeter equation (BSE) 
matrix~\cite{bene+98prl,albr+98prl,rohl-loui98prl} 
are the Gaussian transition elements
$A_{\gamma}^{n{k},n'{k'}}\equiv\langle n {k}
|\chi_\gamma^{{k-k'},h_d}|n'{k'}\rangle$.
They are given by
\begin{equation}
A_{\gamma}^{n{ k},n'{ k'}}
= \sum_{\alpha \beta}
(\psi_{n{ k}}^\alpha)^* \psi_{n'{ k'}}^\beta
M_{\alpha \gamma \beta}^{{ k} h; {k-k'},h_d;
    { k}'h'},
\label{pesante1}
\end{equation}
where $\psi_{n k}^\alpha$ is the Gaussian component
of the wavefunctions of the system [Greek letters 
$\alpha,\beta,\gamma, \ldots$ denote the quartuple 
$(\sigma, n, l, m)$], and $h_d = h(n{ k}) -h(n'{ k'})$ is 
the difference between the quantum number $h$  
for the states $n{ k}$ and $n' { k'}$. 
The
three-basis-function integrals   
$ M_{\alpha \gamma \beta}^{{ k} h; {k-k'},h_d;
    { k}'h'}$ are given by : 
\begin{equation}
M_{\alpha \gamma \beta}^{{ k} h; { k-k'},h_d;
    { k}'h'} = 
\sum_{ii'jj'} 
L_{\alpha\gamma\beta}^{ij;i'j'}
f_{k-k'}^{i}f_{ k'}^{i'} g_{{k-k'},h_d}^{j}g_{{ k}'h'}^{j'} . 
\label{pesante2}
\end{equation}
The phase-multiplied 
three-center integrals  $L_{\alpha\gamma\beta}^{ij;i'j'}$ are 
calculated from the pure three-center integrals
$
I_{ACB}^{{\bf R}{\bf R'}{\bf R''}} = 
\int d^3r \phi^*_A({\bf r}-{\bf R}) 
                                  \phi_C({\bf r}-{\bf R}')
                                  \phi_B({\bf r}-{\bf R}'')
$
with
$L_{\alpha\gamma\beta}^{ij;i'j'}  = k_{m(c)}^{j} 
k_{m(b)}^{j'}I_{ACB}^{\tau ; 
G_2^{j}G_1^{i} \tau'';G_2^{j'}G_1^{i'}\tau'} $
where $k_m = e^{-2\pi i m/N}$. $\tau$, $\tau'$, and $\tau''$ are
the basis vectors (atomic positions in the reduced cell) 
for Gaussians $\alpha$, $\beta$, and $\gamma$.

The evaluation of the
Gaussian transition elements and the three-basis-function 
integrals in Eq.~(\ref{pesante1}) and Eq.~(\ref{pesante2}) is
the most expensive part of the calculation and, for a given
precision,  scales like
$\sim (N N_{ k})^2 \sim R^2$, where $N_{k}$ is the number of $k$-points
and $R$ is the radius of the tube. This scaling is considerably faster
than that of a plane-wave calculation, which scales
like $\sim TR^4 \log(TR^2)$. Our method has the additional advantage of
scaling independently of the cell length $T$ and allows for a  
calculation of tubes like
the (4,2) NT, which have  unusually long cell lengths.

The expressions for the BSE 
matrix are simple functions of the matrix elements in Eq.~(\ref{pesante1}).
We distinguish two cases: (a) If the light is polarized along the 
tube axis, $z$,
the selection rule for $h$ is $h=h'$. For this polarization
we include only those transitions which conserve $h$. In so doing, the
BSE matrix is reduced to $N$ independent blocks.   
(b) The same holds for 
circularly-polarized light with polarization vector 
$\hat{\bf e}_{\pm} =  1/\sqrt{2}(\hat{\bf e}_x \pm i \hat{\bf e}_y)$ with
the only difference that 
the selection rule is $h = h'\pm 1$. The direct term for
both polarizations is:
\begin{equation}
K^d_{cv{ k},c'v'{ k'}} = A_{\alpha}^{ck,c'k'}[ A_{\beta}^{vk,v'k'}]^* 
          W^{\alpha \beta}_{{ k-k'},h(ck)-h(c'k')}.
\label{BSE1}
\end{equation}
The exchange term is:   
\begin{equation}
K^x_{cv{ k},c'v'{ k'}} =  A_{\alpha}^{ck,vk}[ A_{\beta}^{c'k',v'k'}]^* 
          V^{\alpha \beta}_{{ 0},h(ck)-h(vk)} , 
\label{BSE2}
\end{equation}
where $h(ck)-h(vk)=0$~$(\pm 1)$ 
for the $\hat{\bf e}_z$ ($\hat{\bf e}_{\pm}$) polarization,
 $ W^{\alpha \beta}_{{ q},h}$
and $ V^{\alpha \beta}_{{ q},h}$ are 
the screened and bare Coulomb interactions expressed in the
symmetrized basis of Gaussians (the Einstein sum convention 
is used to sum over the Greek indices) calculated 
following the procedure in Ref.~\cite{rohl+95prb}. The screening
is calculated in RPA. Building the matrices in Eq.~(\ref{BSE1}) 
and Eq.~(\ref{BSE2}) does not require much
computation time because the Greek indices range from 1 to 
the number of Gaussians (fifty in this work) which is
small and does not depend on system size.
\begin{figure}
\includegraphics[clip,width=0.98\columnwidth]{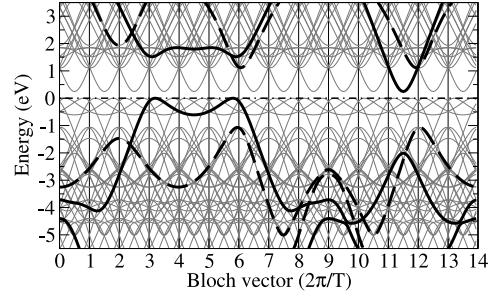}
\caption{\label{bands} Band structure in the extended 
zone scheme. Each 
panel represents a full Brillouin 
zone (FBZ). The solid line represents a 
continuation of a band with 
character $h=1$ in the first Brillouin zone,
and the dashed line begins with a 
band with character $h=2$. The top of the
valence band is 
represent by a horizontal dotted-dashed line.}
\end{figure}

The exciton wavefunction for the $n^{th}$ excitation is:
\begin{equation}
\phi^{(n)}({\bf r}_e,{\bf r}_h) = \sum_{cvk} \Psi^{(n)}_{cvk} 
[\psi_{vk}({\bf r}_h)]^* \psi_{ck}({\bf r}_e), 
\label{eq:excwf} 
\end{equation}
where the expansion coefficients $\Psi^{(n)}_{cvk}$ satisfy the BSE:
\begin{equation}
\![(E_{ck}\!-\!E_{vk}) \delta_{cc'} \delta_{vv'} \delta_{kk'}
\! +\! K_{cvk,c'v'k'}] \Psi^{(n)}_{c'v'k'}\!\! =\! \Omega_n\Psi^{(n)}_{cvk},
\label{eq:BSE}
\end{equation}
with $K_{cvk,c'v'k'} = 2K^x_{cvk,c'v'k'} - K^d_{cvk,c'v'k'}$.
Since we are interested only in the effect of the electron-hole
interaction on the spectra, we use LDA energies for the
$E_{nk}$ of Eq.~\ref{eq:BSE} without quasiparticle corrections~\cite{gw}.
%The single-particle energies $E_{nk}$ appearing in Eq.~\ref{eq:BSE} 
%have been obtained within the local-density approximation,
%without any self-energy correction. Our calculation 
%allows us therefore to predict  
%the difference between the  absorption
%spectra with and without the electron-hole interaction 
%(i.~e.~the exciton binding energy)~\cite{gw}, but cannot predict
%the absolute position of the absorption edge.   
The imaginary part of the dielectric function
is given by
$\epsilon_2(\omega) = {{4 \pi} / \omega^2} 
\sum_n |M_n|^2 \delta(\omega-\Omega_n)$ where
$M_n = \sum_{cvk} \Psi^{(n)}_{cvk} \langle v{\bf k} |{\bf v} \cdot {\bf e}  
 |c{\bf k}  \rangle$.
In this work, the momentum operator is used in lieu of the
velocity operator since the former is easier to express 
in Gaussians. The difference in another carbon-based system,
i.e., diamond, has been shown to be a 15~$\%$ 
to 20~$\%$ effect in the peak intensity of
absorption spectrum~\cite{isma+01prl}, but has a negligible effect
on the peak position.
 
\begin{figure}
\includegraphics[clip,width=0.98\columnwidth]{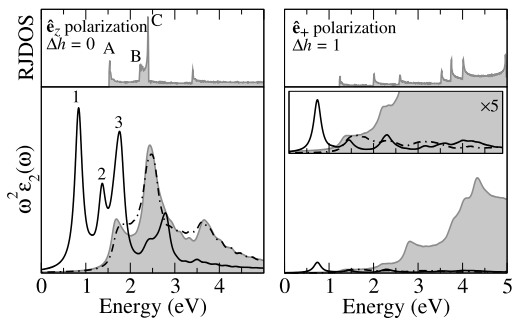}
\caption{\label{abs} The left panels show the RJDOS and
the optical absorption for the $\hat{\bf e}_z$ polarization
and the right panels for the $\hat{\bf e}_{+}$ polarization. In both 
bottom panels,
the solid line represents the spectrum with electron-hole
interaction and the light grey line the spectrum in the single-particle
picture. The dotted-dashed line represents
the spectrum with exchange term only. 
All spectra are computed with a broadening of 0.07~eV,
without self-energy corrections.}  
\end{figure}

We  focus on the carbon (4,2) NT~\cite{foot2}.  
Fig.~\ref{bands} shows its electronic band structure. 
It is represented in the extended zone scheme to
clearly show how the bands unfold and how they are related to the
quantum numbers $h$ and ${\bf k}$. Every panel corresponds to a
FBZ.
The entire band structure is
shown with light grey lines.
The thick solid~(dashed) lines are lines of
constant character $h=1$ ($h=2$).
In this scheme, it is clear
how to represent the set of all optically permitted transitions for both 
polarizations ($\hat{\bf e}_z$ and $\hat{\bf e}_{\pm}$). 
For the $\hat{\bf e}_z$ polarization, 
only the vertical transitions from the solid valence
line to the solid conduction line
 and those from the dashed valence line to the dashed conduction
line are optically allowed. 
For the $\hat{\bf e}_{\pm}$ polarization, 
only the vertical transitions from the solid to dashed line, or viceversa,
depending on the helicity of the polarization, 
are optically allowed. 
Due to symmetry, both 
helicities ( $\hat{\bf e}_{+}$ and $\hat{\bf e}_{-}$)
yield the same spectrum.
In the non-interacting theory, these transitions can be represented
using a reduced joint density of states (RJDOS). It is obtained by 
excluding from the joint density of states all optical transitions
forbidden by symmetry.  
The RJDOS for both polarizations
is shown in the top two panels of Fig.~\ref{abs}.
The corresponding absorption spectra are shown in the bottom two panels.
They are computed
with 24 k-points in the FBZ and
a broadening of 0.07~eV. The BSE matrix computed with this k-point sampling
and for the energy range of interest
is small with dimension $\approx$~300$\times$300.
For the $\hat{\bf e}_z$ polarization, 
the absorption spectrum with excitonic 
effects (solid line of bottom left panel of Fig.~\ref{abs})
 has three peaks: the lowest two peaks correspond to 
bound excitons with binding energies $E_b =0.8$~eV and $E_b =0.2$ eV,
while the third one is an unbound resonance. This
is consistent
with the value of 1.0~eV found for an 
(8,0) nanotube in another work~\cite{spat+04prl}.

\begin{figure}
\includegraphics[clip,width=0.98\columnwidth]{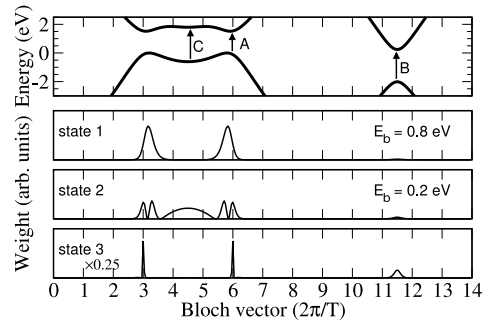}
\caption{\label{transitions} 
The top panel shows a portion of the band structure in the
extended zone scheme with the 
relevant bands [solid bold line ($h=1$) in Fig.~\ref{bands}].
 The bottom three panels show the 
weighted contributions of the transitions which comprise 
three excitonic states that correspond to the first
three peaks in the absorption spectrum for the $\hat{\bf e}_z$
polarization. }
\end{figure}

To gain a better understanding of the nature of the excitonic states
giving rise to these three peaks, we consider the three 
corresponding excitonic states with the highest oscillator
strength $M_n$.
Fig.~\ref{transitions} shows, 
in the extended zone scheme,
the weighted contributions of the optical transitions to
these states (bottom three panels). State 1
is made up of transition A shown in the band diagram (top panel).
State 2 is derived from transitions A and C,  and state 3
from transitions A and B. Transitions A, B, and C correspond
to the Van Hove singularities in the RJDOS labeled in the top left panel
of Fig.~\ref{abs}.
Note that the resonance 
represented by state 3 is coupled with transitions
in the continuum region, 
appearing in Fig.~\ref{transitions} (bottom panel)
as two sharp peaks each corresponding to a single Bloch vector.

From the bottom left panel of Fig.~\ref{abs} one
can directly compare the absorption with both direct and 
exchange terms included in the BSE (solid line),
the absorption with exchange term only (dotted-dashed lines),
and the absorption without excitonic effects (the shaded line): it
is evident that excitonic effects radically alter the
absorption spectrum. Moreover, 
most of the effect comes from the direct term, 
as the exchange term alters the spectrum by a small amount. 

In the bottom right panel of Fig.~\ref{abs}, we represent 
the absorption spectrum for the $\hat{\bf e}_{+}$ polarization:  
the spectrum calculated using the non-interacting theory   
is greatly suppressed when both exchange and direct terms are 
included in the BSE Hamiltonian. This phenomenon was discovered in another
theoretical work on 
nanotubes~\cite{mari+03prl} using time-dependent 
local density approximation~\cite{pete+96prl}. To see if such a 
suppression is to be attributed to the exchange term [equivalent 
to local field effects 
(LFE)], or to the direct electron-hole Coulomb attraction, we
plot an additional 
curve with the exchange term only (dotted-dashed line).
Fig.~\ref{abs} shows that the 
suppression is largely due to the exchange term and
that the direct term affects the spectrum less, in agreement with 
Ref.~\cite{mari+03prl}.
\begin{figure}
\includegraphics[clip,width=0.9\columnwidth]{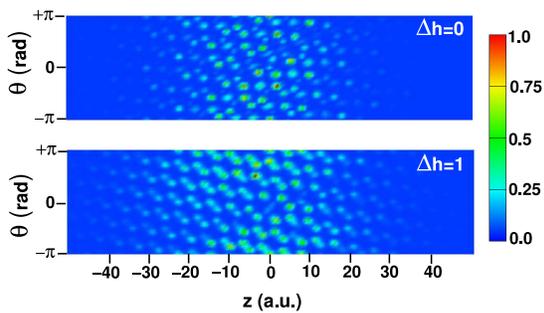}
\caption{\label{excwfc} Exciton wavefunctions for the 
two lowest, optically active excitons of the (4,2) carbon NT.
The top panel shows the $\Delta h = 0$ exciton ($E_b = 0.8$~eV)
and the bottom panel the $\Delta h = 1$ exciton ($E_b = 0.6$~eV).
           The color plots 
           represent the projection on the tube lateral surface 
           of the probability of finding the hole  
           when the electron is fixed in the origin slightly above
           one of the atoms. $z$ is the tube axis.
          $\theta$ is the circumference direction.  }   
\end{figure}
The main effect of the direct term is  
to open up a significant peak
below the lowest van Hove singularity, 
corresponding to a bound, $\Delta h = 1$
exciton with binding energy $E_b = 0.6$~eV.

We plot the exciton wavefunction
$\phi^{(n)}({\bf r}_e,{\bf r}_h)$ [see Eq.~(\ref{eq:excwf})]
 by fixing the position of the electron:    
Fig.~\ref{excwfc} shows the two lowest lying, 
optically active excitons for
the $\hat{\bf e}_z$  ($\Delta h = 0$) 
and the $\hat{\bf e}_{+}$ polarization ($\Delta h = 1$). 
Both excitons are localized 
in the direction of the tube axis, $z$, but delocalized along
the tube circumference, $\theta$. This behavior
is in agreement with the variational calculation of Ref.~\cite{pede03prb}. 
The wavefunction for the even parity (and
hence, optically inactive),
 $\Delta h = 0$
exciton located 0.1~eV below the first absorption peak 
(not shown) is similar
to the lowest optically active (odd parity) exciton.

In conclusion, we have 
developed an efficient method for performing
BSE calculations on NTs of all sizes and chiralities
that uses a local, symmetry-based approach.
Applying this method to a (4,2) NT, 
we have shown that excitonic effects are important 
and have quantitatively determined these effects. 
For the $\hat{\bf e}_z$ polarization ($\Delta h = 0$), 
the excitonic effects give rise
to three enhanced peaks, and the lowest excitation is 
 0.8~eV below the onset of the single-particle continuum. 
 The corresponding 
exciton is  localized along  ${\bf \hat{e}}_z$, but
delocalized along the tube circumference.
For the $\hat{\bf e}_{+}$ polarization($\Delta h = 1$), 
the exchange part of the BSE matrix drastically
suppresses the absorption spectrum. The direct part opens up a large
peak 0.6~eV below the single-particle singularity 
corresponding to a bound, localized exciton.

We are grateful to  G. Goldoni,  J. Menendez, and M. Rohlfing  
for fruitful discussion.  Computer time was partly provided 
by CINECA through INFM Parallel Computing Projects.  
The support by the RTN EU Contract ``EXCITING'' No.  HPRN-CT-2002-00317,
and by FIRB ``NOMADE'' is also acknowledged.


\begin{thebibliography}{10}
\bibitem{foot1}
Typical values of exciton binding energies are of the order of 0.01-0.1 eV in
  semiconductor wires~\cite{ross-moli96prl} and 0.1-1 eV in polymers
  ~\cite{ruin+02prl}.
\bibitem{gw}
We have calculated
the self energy. The quasiparticle correction increases the
  valence band width by 17~$\%$,  decreasing in turn  
  $E_b$ by only
  0.03~eV.
\bibitem{foot2}
In this work, we have used a cylindrical, unrelaxed structure with a C-C bond
  length of 1.42 $\AA$.
\end{thebibliography}
\end{document}